\documentclass[paper, nofootinbib]{revtex4}
\usepackage{amsfonts}
\usepackage{graphicx}
\usepackage[latin1]{inputenc}
\usepackage{amsmath}
\usepackage{amssymb}

\begin{document}


\title{Relativistic corrections to the algebra of position variables and spin-orbital interaction
}
\author{Alexei A. Deriglazov}
\email{alexei.deriglazov@ufjf.edu.br}
\affiliation{%
Depto. de Matem\'atica, ICE, Universidade Federal de Juiz de Fora, MG, Brasil \\ and \\ Laboratory of Mathematical
Physics, Tomsk Polytechnic University, 634050 Tomsk, Lenin Ave. 30, Russian Federation}

\author{Andrey M. Pupasov-Maksimov}
 \email{pupasov@phys.tsu.ru}
\affiliation{%
Depto. de Matem\'atica, ICE,\\
Universidade Federal de Juiz de Fora, MG, Brasil
}%
\date{\today}

\begin{abstract}
In the framework of vector model of spin, we discuss the problem of a covariant formalism \cite{Pomeranskii1998}
concerning the discrepancy between relativistic and Pauli Hamiltonians. We show how the spin-induced non commutativity
of a position accounts the discrepancy on the classical level, without appeal to the Dirac equation and
Foldy-Wouthuysen transformation.
\end{abstract}

\keywords{Semiclassical Description of Relativistic Spin, Dirac Equation, Theories with Constraints}

\maketitle

\section{Introduction \label{intro}}

In series of previous works \cite{deriglazov-bmt:2013,DPM3,DPM1,deriglazov2014Monster}  we developed a
Poincare-invariant variational formulation describing  particle with spin. This classical model provides a unified
description of both Frenkel and BMT equations \cite{bmt:59}. The latter are considered as a basic tool in the analysis
of the polarization precession measurements \cite{miller2007muon}. In \cite{DPW2} we extended the variational
formulation to the general relativity, where the classical models of a spinning particle are widely used to describe a
rotating body in pole-dipole approximation
\cite{dixon:1964,Khriplovich1989,Khriplovich2008,alwal2015_3,alwal2015_2,alwal2015_1,Koch2015,Balakin2015,Frob2016}.
Another possible application can be related with the kinetic theory of chiral medium, where, in the regime of weak
external fields and weak interactions between spinning (quasi)-particles, each particle can be considered as moving
along a classical trajectory \cite{chen2014}.

For variational formulations provide a starting point to the canonical quantization \cite{dirac1958}, they have
incredible theoretical importance connecting classical and quantum descriptions of nature. Canonical quantization of
the free spinning particle (within our variational formulation, \cite{deriglazov2013quant}) leads to the one-particle
relativistic quantum mechanics with positive-energy states. It also identifies the non-commutative Pryce's d-type
center of mass operator \cite{deriglazov2013quant} \footnote{See also \cite{gitman1990quantization}, where the same
result was obtained for the classical particle with anticommuting spin variables.} as the quantum observable which
corresponds to the classical position variable. The noncommutative position variables were constructed already by Pryce
\cite{pryce1948mass}. He shown that coordinates of the relativistic center-of-mass have to obey non-trivial Poisson
brackets. As a result, the corresponding quantum observables do not commute. Therefore a physically meaningful position
operators of a spin-$1/2$ particle should be non-commutative.

Recent theoretical studies revive Snyder's attempts \cite{snyder1947quantized} to solve fundamental physical problems
by introducing non-commutativity of the space \cite{deriglazov-nc:2003}. It is believed that this fundamental
non-commutativity may be important at Plank length scale $\lambda_P$. Extensive studies of non-commutativity cover both
classical and quantum theories, as well relativistic and non-relativistic situations. Postulating non-commutative
deformation of position operators \cite{Ferrari:2012bv} one can study physical consequences and estimate possible
effects. Calculations of the hydrogen spectrum corrections strongly limit possible non-commutativity of coordinate
parameters in the Dirac equation
\cite{sheikh2001,Gomes2010spin-non-commutativity,khodja2012ncommutative-coulomb,santos2012zeeman-non-commutative,kupriyanov2013hydrogen-non-commutativity}.

In the present work we study effects of a non-commutativity of Pryce's d-type center of mass (at both classical and
quantum levels) in the description of electron interacting with an electromagnetic background. Our considerations
extend results of \cite{deriglazov2013quant} towards a quantization of interacting spinning particle.

In the free theory, different candidates for the position operator are almost indistinguishable. All these operators
obey the same Heisenberg equations (uniform rectilinear motion), and the difference in their expectation values is of
Compton wave-length order, $\lambda_C$. In the interacting case, the problem of the identification of quantum position
observables becomes more complicated.\footnote{ Another related problem is in the definition of spin operator, since it
correlates with the definition of a center of mass. Bauke in \cite{bauke2014spin} compares Pauli, Foldy-Wouthuysen,
Czachor, Frenkel, Chakrabarti, Pryce, and Fradkin-Good spin operators in different physical situations and concluded
that interaction with electromagnetic potentials allows to distinguish between various spin operators experimentally. }
Fleming \cite{fleming1965nonlocal} noted: {\it "The simplest form of interaction is that due to a static potential
which may be expressed in terms of the position operator of the particle. For a relativistic particle, however, the
important question arises of which position operator should be used. The conventional approach, in which the position
operator is assumed to be local, forces the choice of the center of spin."}\footnote{Fleming calls the Newton-Wigner
position operator as the center of spin, while Pryce d-type operator is called as the center of mass.} He also
observed, that a formal substitution of Pryce d-type operator into the potential leads to some reasonable corrections:
{\it "The first correction term to a spherically symmetric local potential will be recognized as the spin-orbit
coupling that Thomas derived many years ago as a consequence of classical relativity and which appears in the
nonrelativistic limit of the Dirac equation for spin particles."}

Analogous situation was observed in general relativity,
\cite{feynman1961,fleming1965covariant,corben:1968,Pomeranskii1998} where a formal substitution of a non-local position
variable into potential results in correct equations of motion for the spinning particle. Restricting ourselves to the
case of special relativity, in the present work we provide some theoretical grounds for such substitution.

The paper is organized as follows. In Sect. 2 we formulate the problem of covariant formalism \cite{Pomeranskii1998}
concerning the discrepancy between relativistic and Pauli Hamiltonians. In Sect. 3 we give a brief description of the
vector model for the classical description of a relativistic spinning particle. In Sect. 4 we consider canonical
quantization of the model in the physical-time parametrization and realize classical algebra of Dirac brackets by
quantum operators in the case of a stationary electro-magnetic background. This realization deforms the covariant
Hamiltonian and at low energies gives Pauli Hamiltonian with correct spin-orbital interaction. In the conclusion we
discuss the obtained results.

\section{Model independent discussion of the quantum and classical Hamiltonians of a spinning particle}
From quantum point of view, at low energies an electron interacting with a background electromagnetic field is
described by the two-component Schr\"{o}dinder equation. Pauli Hamiltonian\footnote{We will write quantum Hamiltonians
and other operators using the hat, the same observables without the hat correspond to the classical theory. Thus
(\ref{pauli-hamiltonian}) defines also classical Pauli-like Hamiltonian.} includes spin-orbital and Zeeman interactions
\begin{eqnarray}
\hat H_{ph}=\frac{1}{2m}(\hat{{\rm {\bf p}}} - \frac{e}{c}{\rm {\bf A}})^2 - eA_0 + \frac{e(g-1)}{2m^2c^2}\hat{{\rm
{\bf S}}}[\hat{\bf p}\times{\bf E}]- \frac{eg}{2mc}{\rm {\bf B}}\hat{{\rm {\bf S}}} =\hat H_{charge}+\hat
H_{spin-em}\,, \qquad \qquad \label{pauli-hamiltonian}
\end{eqnarray}
where the spin operator is proportional to $\sigma$-matrices of Pauli $\hat S^i=\frac{\hbar}{2}\sigma^i$. Gyromagnetic
ratio $g$ is a coupling constant of spin with an electromagnetic field. In principle, in non-relativistic theory one
can expect different coupling constants for the third and the fourth terms of the Hamiltonian. Experimental
observations of the hydrogen spectrum lead to the factor $g-1$ in the third term and to the factor $g$ in the last
term. Thus, the Hamiltonian explains Zeeman effect and reproduces fine structure of the energy levels of the hydrogen
atom. This Hamiltonian follows also from the non-relativistic limit of the Dirac equation in the Foldy-Wouthuysen
representation \cite{dirac1958,foldy:1978}.

From classical point of view, models of spinning particles are based on a Lagrangian or Hamiltonian mechanics, both in
the relativistic and non-relativistic regime \cite{AAD2010}. In a covariant formulation, the spin part of the
Hamiltonian describing an interaction between spin $S$ and electromagnetic field reads
\begin{eqnarray}
H_{spin-em-cov}\sim\frac{eg}{2m^2c^2}{\rm {\bf S}}[{\bf p}\times{\bf E}]- \frac{eg}{2mc}{\rm {\bf B}}{\rm {\bf S}}\,.
\label{FF0.2.1}
\end{eqnarray}
We emphasize that the expression (\ref{FF0.2.1})  follows from the analysis of all possible terms in covariant
equations of motion and thus is a model-independent \cite{Pomeranskii1998}. It can also be predicted from symmetry
considerations on the level of a Hamiltonian. For instance, if we take the Frenkel spin-tensor $S^{\mu\nu}$ with the
covariant condition $S^{\mu\nu}P_\nu=0$, the only Lorentz-invariant combination that could give the desired terms
written in (\ref{FF0.2.1}) is $F_{\mu\nu}S^{\mu\nu}=2E^iS^{i0}+\epsilon^{ijk}S^{ij}B^k$ (see our notations in
Appendix).

For the classical gyromagnetic ratio $g=2$, the classical spin-orbital interaction in (\ref{FF0.2.1}) differs by the
famous and troublesome factor\footnote{This factor is often referred to Thomas precession \cite{thomas1926motion}. We
will not touch this delicate and controversial issue \cite{frenkel:1926,stepanov2012} since the covariant formalism
automatically accounts the Thomas precession \cite {weinbergGC}.} of $\frac12$ from its quantum counterpart in
(\ref{pauli-hamiltonian}). It seems that quantization of  $H_{spin-em-cov}$ will not reproduce quantum behavior given
by $\hat H_{spin-em}$. The issue about this difference was raised already in 1926 \cite{frenkel:1926} and still remains
under discussion \cite{Pomeranskii1998}.

In principle, Hamiltonian $H_{spin-em}$ can be obtained, if one impose a non covariant supplementary condition on spin,
$2S^{i0}P_0+S^{ij}P_j=0$, where $P_0\sim -mc$ in the leading approximation. On a first glance, any covariant
spin-supplementary condition \cite{frenkel:1926,Papapetrou:1951pa,pirani:1956,dixon:1964,tulczyjew:1959} would give
$H_{spin-em-cov}$ and the discrepancy factor of $\frac12$.

In the next section we study this issue in the framework of vector model of a spinning particle
\cite{deriglazov2014Monster}. We show that the vector model provides an answer on a pure classical ground, without
appeal to the Dirac equation. In a few words, it can be described as follows. The relativistic vector model involves a
second-class constraints, which should be taken into account by passing from the Poisson to Dirac bracket. The
emergence of a higher non linear classical brackets that accompany the relativistic Hamiltonian (\ref{FF0.2.1}) is a
novel point, which apparently has not been taken into account in literature. If we pretend to quantize the model, it is
desirable to find a set of variables with the canonical brackets. The relativistic Hamiltonian (\ref{FF0.2.1}), when
written in the canonical variables, just gives (\ref{pauli-hamiltonian}).

\section{Vector model of spinning particle in the parametrization of physical time.}

To find the classical brackets that accompany $H_{cov}$ we need a systematically developed model of a spinning
particle. Here we consider the vector model and briefly describe the construction of the Hamiltonian and the brackets
in a stationary electromagnetic field. For a detailed discussion of the model, see \cite{deriglazov2014Monster}.

Configuration space of the vector model of spinning particle is parameterized by a point $x^\mu(\tau)$ of a world-line
and a vector $\omega^\mu(\tau)$ attached to that point. The configuration-space variables are taken in an arbitrary
parametrization $\tau$ of the world-line. The conjugate momenta of the variables are denoted by $p^\mu$ and $\pi^\mu$,
correspondingly. Frenkel spin-tensor in the vector model is a composite quantity,
$S^{\mu\nu}=2(\omega^\mu\pi^\nu-\omega^\nu\pi^\mu)$. The free Lagrangian can be written in a number of equivalent forms
\cite{alwal2015_3,deriglazov2013quant}. To describe the spin-field interaction through the gyromagnetic ratio $g$, we
use the Lagrangian with an auxiliary variable $\lambda(\tau)$
\begin{eqnarray}\label{m.1}
S=\int d\tau\frac{1}{4\lambda}\left[\dot xN\dot x+D\omega ND\omega- \sqrt{\left[\dot xN\dot x+D\omega
ND\omega\right]^2-4(\dot xND\omega)^2}\right]- \cr \frac{\lambda}{2}(m^2c^2-\frac{\alpha}{\omega^2})+\frac{e}{c}A\dot
x, \qquad \qquad \qquad
\end{eqnarray}
where
$D\omega^\mu\equiv\dot\omega^\mu-\lambda\frac{eg}{2c}F^{\mu\nu}\omega_\nu$.
The auxiliary variable provides a homogeneous transformation law of $D\omega$ under the reparametrizations,
$D_{\tau'}\omega=\frac{d\tau}{d\tau'}D\omega$. The matrix $N_{\mu\nu}$ is the projector on the plane orthogonal to
$\omega^\nu$, $N_{\mu\nu}= \eta_{\mu\nu}-\frac{\omega_\mu \omega_\nu}{\omega^2}$. The parameter $m$ is mass, while
$\alpha$ determines the value of spin. The value $\alpha=\frac{3\hbar^2}{4}$ is fixed by quantization conditions and
corresponds to an elementary spin one-half particle. In the spinless limit, $\alpha=0$ and $\omega^\mu=0$, the
functional (\ref{m.1}) reduces to the well known Lagrangian of the relativistic particle, $\frac{1}{2\lambda}\dot
x^2-\frac{\lambda}{2}m^2c^2+\frac{e}{c}A\dot x$.

Frenkel considered the case $g=2$ and found approximate equations of motion neglecting quadratic and higher terms in
spin, fields and field gradients. Equations of motion obtained from (\ref{m.1}) coincide with those of Frenkel in this
approximation \cite{frenkel:1926}.

To find relativistic Hamiltonian in the physical-time parametrization\footnote{Which is necessary for the canonical
quantization.}, we use the Hamiltonian action associated with (\ref{m.1}). This reads \cite{deriglazov2014Monster},
$\int d\tau ~ p\dot x+\pi\dot\omega-\lambda_iT_i$, where $\lambda_i$ are Lagrangian multipliers associated with the
primary constraints $T_i$. The variational problem provides both equations of motion and constraints of the vector
model in an arbitrary parametrization. Using the reparametrization invariance of the functional, we take physical time
as the evolution parameter, $\tau=\frac{x^0}{c}=t$, then the functional reads
\begin{eqnarray}\label{ch09:eqn9.7.1}
S_H=\int dt ~  c\tilde{\cal P}_0-eA^0+p_i\dot x^i+\pi_\mu\omega^\mu-  \qquad \qquad \qquad \cr
\left[\frac{\lambda}{2}\left(-\tilde{\cal P}_0^2+{\cal
P}_i^2-\frac{eg}{4c}(FS)+m^2c^2+\pi^2-\frac{\alpha}{\omega^2}\right)+ \lambda_2\omega\pi+\lambda_3{\cal
P}\omega+\lambda_4{\cal P}\pi\right],
\end{eqnarray}
where  $\tilde{\cal P}_0=p_0-\frac{e}{c}A_0$ and ${\cal P}^i=p^i-\frac{e}{c}A^i$ is $U(1)$\,-invariant canonical
momentum.

We can treat the term associated with $\lambda$ as a kinematic constraint of the variational problem. Following the
known prescription of classical mechanics, we solve the constraint,
\begin{eqnarray}\label{ch09:eqn9.7.2}
\tilde{\cal P}_0=-\tilde{\cal P}^0=-\sqrt{{\cal P}_i^2-\frac{eg}{4c}(FS)+m^2c^2+\pi^2-\frac{\alpha}{\omega^2}},
\end{eqnarray}
and substitute the result back into Eq. (\ref{ch09:eqn9.7.1}), this gives an equivalent form of the functional
\begin{eqnarray}\label{ch09:eqn9.7.3}
S_H=\int dt ~  p_i\dot x^i+\pi_\mu\dot\omega^\mu- \left[c\sqrt{{\cal P}_i^2-\frac{eg}{4c}(FS)+
m^2c^2+\pi^2-\frac{\alpha}{\omega^2}}+eA^0+  \right. \cr \left. \lambda_2\omega_\mu\pi^\mu+\lambda_3{\cal
P}_\mu\omega^\mu+\lambda_4{\cal P}_\mu\pi^\mu\right], \qquad \qquad \qquad
\end{eqnarray}
where the substitution (\ref{ch09:eqn9.7.2}) is implied in the last two terms as well. The sign in front of the square
root (\ref{ch09:eqn9.7.2}) was chosen according to the right spinless limit, $L=-mc\sqrt{-\dot x^\mu\dot x_\mu}$. The
expression in square brackets is the Hamiltonian.

The variational problem implies the first-class constraints $T_2\equiv\omega\pi=0$,
$T_5\equiv\pi^2-\frac{\alpha}{\omega^2}=0$. They determine gauge symmetries and physical observables of the theory. The
quantities $x^i(t)$, ${\cal P}^i(t)$ and $S^{\mu\nu}(t)$ have vanishing Poisson brackets with the constraints and hence
are candidates for observables.  The set
\begin{eqnarray}\label{ch09:eqn9.7.4}
T_3=-{\cal P}^0\omega^0+{\cal P}^i\omega^i=0, \qquad T_4=-{\cal P}^0\pi^0+{\cal P}^i\pi^i=0,
\end{eqnarray}
where
\begin{eqnarray}\label{ch09:eqn9.7.5}
{\cal P}^0\equiv\sqrt{{\cal P}_i^2-\frac{eg}{4c}(FS)+m^2c^2}.
\end{eqnarray}
represents a pair of second class constraints. In all expressions below the symbol ${\cal P}^0$ represents the function
(\ref{ch09:eqn9.7.5}). The constraints imply the spin-supplementary condition
\begin{eqnarray}\label{ssc}
S^{\mu\nu}{\cal P}_\nu=0,
\end{eqnarray}
as well as the value-of-spin condition $S^{\mu\nu}S_{\mu\nu}=8\alpha$.

To represent the Hamiltonian in a more familiar form, we take into account the second-class constraints by passing from
Poisson to Dirac bracket. As the constraints involve conjugate momenta of the position ${\bf x}$, this leads to
nonvanishing brackets for the position variables. In the result, the position space is endowed, in a natural way, with
a noncommutative structure which originates from accounting of spin degrees of freedom. For the convenience, an exact
form of Dirac brackets of our observables is presented in the Appendix. Since the Dirac bracket of any quantity with
second-class constraints vanishes, we can omit them from the Hamiltonian. The first-class constraints can be omitted as
well, as they do not contribute into equations of motion for physical variables. In the result we obtain the
relativistic Hamiltonian
\begin{eqnarray}\label{pht.16}
H_{cov}=c\sqrt{\vec{\cal P}^2-\frac{eg}{4c}F_{\mu\nu}S^{\mu\nu}+m^2c^2}+eA^0.
\end{eqnarray}
Note that the Dirac brackets (\ref{pht14.2}), (\ref{pht15.1}) encode the most part of spin-field interaction, on this
reason we have arrived at a rather simple form for the physical Hamiltonian. Equations of motion follow from this
Hamiltonian with use of the Dirac bracket\footnote{We emphasize that the use of canonical brackets will lead to
different equations. In our opinion, this turns out to be the reason for debates around the controversial results
obtained by different groups, see the discussion in \cite{Pomeranskii1998}.}: $\frac{d z}{dt}=\{z, H_{cov}\}_D$.

\section{First Relativistic Corrections and Fine Structure of Hydrogen Spectrum}
\label{ch09:sec9.7}

To quantize our relativistic theory we need to find quantum realization of highly non linear classical brackets
(\ref{pht14.2}). They remain non canonical even in absence of interaction. For instance, the first equation from
(\ref{pht14.2}) in a free theory reads $\{x^i, x^j\}=\frac{1}{2mcp^0}S^{ij}$. It is worth noting that non relativistic
spinning particle \cite{AAD2010,DPM1} implies the canonical brackets, so the deformation arises as a relativistic
correction induced by spin of a particle. Technically, the deformation arises from the fact that the constraints
(\ref{ch09:eqn9.7.4}), used to construct the Dirac bracket, mixes up the space-time and inner-spin coordinates.
Concerning quantum realization of the brackets in a free theory and relativistic covariance of the resulting quantum
mechanics, see \cite{deriglazov2013quant}. In an interacting theory, the explicit form of the brackets is not known.
Therefore we quantize the interacting theory perturbatively, considering $c^{-1}$ as a small parameter and expanding
all quantities in power series. Let us consider the approximation $o(c^{-2})$ neglecting $c^{-3}$ and higher order
terms. For the Hamiltonian (\ref{pht.16}) we have $H_{cov}\approx mc^2+\frac{\boldsymbol{\cal
P}^2}{2m}-\frac{\boldsymbol{\cal P}^4}{8m^3c^2}-\frac{eg}{8mc}(FS)$. Since the last term is of order $(mc)^{-1}$,
resolving the constraint $S^{\mu\nu}{\cal P}_\nu=0$ with respect to $S^{i0}$ we can approximate ${\cal P}^0=mc$, then
$S^{i0}=\frac{1}{mc}S^{ij}{\cal P}^j$. Using this expression we obtain
\begin{eqnarray}\label{pht.16.0}
H_{cov}= mc^2+\frac{\boldsymbol{\cal P}^2}{2m}-\frac{\boldsymbol{\cal P}^4}{8m^3c^2}+eA^0+ \frac{eg}{2mc}
\left[\frac{1}{mc}{\rm {\bf S}}[{\boldsymbol{\cal P}}\times{\bf E}]- {\rm {\bf B}}{\rm {\bf
S}}\right]+o\left(\frac{1}{c^2}\right)\cr =H_{charge}+H_{spin-em-cov}+o\left(\frac{1}{c^2}\right). \qquad \qquad
\end{eqnarray}
Due to the second and fourth terms, we need to know the operators $\hat{\cal P}^i$ and $\hat x^i$ up to order $c^{-2}$,
while $\hat S^{ij}\sim\hat{\bf S}$ should be found up to order $c^{-1}$. With this approximation, the commutators
$[\hat x, \hat x]$, $[\hat x, \hat{\cal P}]$, and $[\hat{\cal P}, \hat{\cal P}]$ can be computed up to order $c^{-2}$,
while the remaining commutators can be written only up to $c^{-1}$. Therefore,  we expand the right hand sides of Dirac
brackets (\ref{pht14.2}) in this approximation
\begin{eqnarray}\label{16.1}
\{x^i, x^j\} & = & \frac{1}{2m^2c^2}S^{ij}+o\left(\frac{1}{c^2}\right), \nonumber \\
\label{16.2}
\{x^i, {\cal P}^j\} & = & \delta^{ij}+o\left(\frac{1}{c^2}\right), \nonumber  \\
\label{16.3}
\{x^i, S^{jk}\} &= & 0+o\left(\frac{1}{c}\right), \\
\label{16.4} \{{\cal P}^i, {\cal P}^j\} &= &\frac{e}{c} F^{ij}+o\left(\frac{1}{c^3}\right), \nonumber \\
\label{16.5} \{{\cal P}^i, S^{jk}\} &= & o\left(\frac{1}{c^2}\right),  \nonumber \\
\label{16.6} \{S^{ij}, S^{kl}\} &= &
2(\delta^{ik}S^{jl}-\delta^{il}S^{jk}-\delta^{jk}S^{il}+\delta^{jl}S^{ik})+o\left(\frac{1}{c}\right). \nonumber
\end{eqnarray}
Only the first bracket acquires a non standard form in the leading approximation. An operator realization of these
brackets on the space two-component Weyl spinors reads
\begin{eqnarray}
\hat{\cal P}_i &=& -i\hbar\frac{\partial}{\partial x^i}-\frac{e}{c}{A}_i({\bf x}),  \label{16.7} \\
\hat{x}_i & =& x_i-\frac{\hbar}{4m^2c^2}\epsilon_{ijk}\hat{P}^j\sigma^k,  \label{16.8} \\
\hat{S}^{ij}& = &\hbar\epsilon_{ijk}\sigma^k,  \label{16.9}  \qquad
\end{eqnarray}
then
\begin{eqnarray}
\hat S^i & = & \frac14\epsilon_{ijk}S^{jk}=\frac{\hbar}{2}\sigma^i, \label{pht.16.9} \\
\hat{S}^{i0} & =& \frac{\hbar}{mc}\epsilon_{ijk}\hat{\cal P}^j\sigma^k. \label{pht.16.10}
\end{eqnarray}
By construction of a Dirac bracket, the operator $\hat S^{i0}$ automatically obeys the desired commutators up to order
$c^{-1}$. The operator $\hat x_i$ coincides with the positive-energy part of Pryce (d) operator in the Foldy-Wouthuysen
representation, see \cite{deriglazov2013quant} for details.

We substitute these operators into the classical Hamiltonian (\ref{pht.16.0}). Expanding $A^0(\hat{\bf x})$ in a power
series, we obtain an additional contribution of order $c^{-2}$ to the potential due to non commutativity of the
position operator
\begin{eqnarray}\label{17}
eA^0\left(x_i-(2mc)^{-2}\epsilon_{ijk}\hat{\cal P}^j\hat S^k\right) \approx eA^0({\bf x})-\frac{e}{2m^2c^2}\hat{\bf
S}[\hat{\boldsymbol{\cal P}}\times{\bf E}].
\end{eqnarray}
The contribution has the same structure as fifth term in the Hamiltonian (\ref{pht.16}).  In the result, the quantum
Hamiltonian up to order $c^{-2}$ reads
\begin{equation}\label{18}
\hat H_{cov}= mc^2+\frac{\hat{\boldsymbol{\cal P}}^2}{2m}-\frac{\hat{\boldsymbol{\cal
P}}^4}{8m^3c^2}+eA^0+\frac{e(g-1)}{2m^2c^2} \hat{\bf S}[\hat{\boldsymbol{\cal P}}\times{\bf E}]-\frac{eg}{2mc}{\bf
B}\hat{\bf S}.
\end{equation}
The first three terms corresponds to an increase of relativistic mass. The last two terms coincides with those in Eq.
(\ref{pauli-hamiltonian}). We could carry out the same reasoning in the classical theory, by asking on the new
variables $z'$ that obey the canonical brackets as a consequence of Eq. (\ref{16.3}). In the desired approximation they
are ${\cal P}^i={\cal P}'^i-\frac{e}{c}A^i(x'^j)$, $x^i=x'^i-\frac{1}{4m^2c^2}S'^{ij}{\cal P}'^j$ and $S^{ij}=S'^{ij}$.
In the result, we have shown that non-commutativity of electron's position  at the Compton scale is responsible for the
fine structure of hydrogen atom.

\section{Conclusions}

Vector model of relativistic spinning particle gives an example of a noncommutative system. In the leading
approximation, the non commutative geometry induced by spin affects only the brackets of position variables, see the
first equation from (\ref{16.1}). The ``parameter of noncommutativity'' is being proportional to spin-tensor. As a
consequence, canonical quantization of the model in leading approximation gives the Pauli Hamiltonian.
Our calculations show that\\
1) classical interaction of spin with electromagnetic field can be described by manifestly covariant term
$S^{\mu\nu}F_{\mu\nu}$, accompanied by the covariant spin-supplementary condition $S^{\mu\nu}P_\nu=0$;\\
2) phase space is endowed with a non-trivial symplectic structure (Dirac brackets), in particular, position variables
become non-commutative due to non-vanishing Dirac brackets;\\
3) the Thomas precession automatically
appears in the equations of motion \cite{deriglazov2014Monster} due to non-trivial Dirac bracket, without modification of the Hamiltonian;\\
4) quantization of the vector model for free electron leads to one-particle relativistic quantum mechanics with
positive-energy states. The free Hamiltonian acts in the space of two-component spinors and reads $\hat
H_{phys}=\sqrt{\hat{\bf p}{\,}^2+m^2c^2}$,
position operator of a free electron is the Pryce's d-type \cite{pryce1948mass,deriglazov2013quant};\\
5) quantization of the model in the case of a stationary electromagnetic background formally leads to the Hamiltonian
$${\hat H_{cov}(F)=
c\sqrt{(\hat{\bf p}{\,}-\frac{e}{c}{\bf A}(\hat{\bf x}))^2 -\frac{eg}{4c}\hat{S}^{\mu\nu}F_{\mu\nu}(\hat{\bf
x})+m^2c^2}+eA^0(\hat{\bf x})}\,,$$ which, up to $o(c^{-2})$ order, coincides with the positive energy part of
Dirac Hamiltonian in the Foldy-Wouthuysen representation. It would be interesting to compare high-order terms;\\
6) non-commutativity of position operator results in the Thomas $1/2$-correction of spin-orbital interaction coming
from $eA^0(\hat{\bf x})$ term\footnote{Similar corrections were obtained in
\cite{kupriyanov2013hydrogen-non-commutativity}. However, they appear from the non-commutativity introduced in the
Dirac representation, therefore they give additional contribution to the correct spectrum as if non-commutativity acts
twice.}.

In the considered approximation our Hamiltonian $\hat H_{phys}(F)$ coincides with the Pauli Hamiltonian for the case of
stationary fields. Therefore, within this approximation there is no any difference between standard and non-commutative
approach to the spin-orbital interaction except a conceptual one. However, in the case of non-stationary fields the
classical Hamiltonian changes form. Further studies of time-dependent electromagnetic fields and next order corrections
may give suggestions for the experimental searches of effects produced by non-commutativity.

{\bf Acknowledgments} The work of AAD has been supported by the Brazilian foundations CNPq (Conselho Nacional de
Desenvolvimento Cientifico e Tecnol\'{o}gico - Brasil) and FAPEMIG (Funda\c{c}\~{a}o de Amparo \`{a} Pesquisa do Estado
de Minas Gerais - Brasil). The work of AMPM is suported by FAPEMIG (Demanda Universal 2015).


\section{Appendix}

{\bf Notation.}  Our variables are taken in arbitrary parametrization $\tau$, then $\dot x^\mu=\frac{dx^\mu}{d\tau}$.
The square brackets mean antisymmetrization, $\omega^{[\mu}\pi^{\nu]}=\omega^\mu\pi^\nu-\omega^\nu\pi^\mu$. For the
four-dimensional quantities we suppress the contracted indexes and use the notation $\dot x^\mu N_{\mu\nu}\dot
x^\nu=\dot xN\dot x$,  $N^\mu{}_\nu\dot x^\nu=(N\dot x)^\mu$, $\omega^2=\eta_{\mu\nu}\omega^\mu\omega^\nu$,
$\eta_{\mu\nu}=(-, +, +, +)$, $\mu=(0, i)$, $i=1, 2, 3$,  Notation for the scalar functions constructed from
second-rank tensors are $FS= F_{\mu\nu}S^{\mu\nu}$, $S^2=S^{\mu\nu}S_{\mu\nu}$.

Electromagnetic field:
\begin{eqnarray}\label{L.0}
F_{\mu\nu}=\partial_\mu A_\nu-\partial_\nu A_\mu=(F_{0i}=-E_i, ~ F_{ij}= \epsilon_{ijk}B_k), \cr
E_i=-\frac{1}{c}\partial_tA_i+\partial_i A_0, \quad B_i=\frac12\epsilon_{ijk}F_{jk}=\epsilon_{ijk}\partial_j A_k.
\nonumber
\end{eqnarray}
Spin-tensor:
\begin{eqnarray}\label{L.0.1}
S^{\mu\nu}=2(\omega^\mu\pi^\nu-\omega^\nu\pi^\mu)=(S^{i0}=D^i, ~ S_{ij}=2\epsilon_{ijk}S_k), \nonumber
\end{eqnarray}
then $S_i=\epsilon_{ijk}\omega_j\pi_k=\frac{1}{4}\epsilon_{ijk}S_{jk}$. Here $S_i$ is three-dimensional spin-vector of
Frenkel and $D_i$ is dipole electric moment.

{\bf Dirac bracket.} Dirac bracket for the constraints (\ref{ch09:eqn9.7.4}) reads
\begin{eqnarray}\label{pht.13}
\{A, B\}_{D}=\{A, B\}-\{A, T_3\}\{T_4, T_3\}^{-1}\{T_4, B\}- \{A, T_4\}\{T_3, T_4\}^{-1}\{T_3, B\}. \qquad \qquad
\nonumber
\end{eqnarray}
Complete list of brackets computed in an arbitrary parametrization can be found in \cite{DPM3}. Here we present the
brackets of the observables $x^i(t)$, ${\cal P}^i(t)$ and $S^{\mu\nu}(t)$. To compute them, we use the auxiliary
Poisson brackets shown in the table \ref{tabular:poissonbr2}.
\begin{table}
\caption{Auxiliary Poisson brackets} \label{tabular:poissonbr2}
\begin{center}
\begin{tabular}{c|c|c|c}
${}$                                & $\{{\cal P}^0, *\}$                     & $\{T_3, *\}$             & $\{T_4, *\}$   \\

\hline \hline

$x^i$       & $-\frac{{\cal P}^i}{{\cal P}^0}$    & $-\omega^i+\frac{\omega^0{\cal P}^i}{{\cal P}^0}$    & $-\pi^i+\frac{\pi^0{\cal P}^i}{{\cal P}^0}$ \\

\hline

$\mathcal{P}^i$  & $-\frac{e}{{\cal P}^0c}[(F\vec {\cal P})^i+\frac{g}{8}\partial^i(SF)]$      &
$\frac{e\omega^0}{{\cal P}^0c}[(F\vec {\cal P})^i+\frac{}{8}\partial^i(SF)]-$       & $\frac{e\pi^0}{{\cal
P}^0c}[(F\vec
{\cal P})^i+\frac{g}{8}\partial^i(SF)]-$ \\

  &  & $\frac{e}{c}(F\vec\omega)^i$   &  $\frac{e}{c}(F\vec\pi)^i$  \\

\hline

${\cal P}^0$  & $0$ & $\frac{e}{2{\cal P}^0c}[(g-2)(\vec{\cal P}F\vec\omega)+$
& $\frac{e}{2{\cal P}^0c}[(g-2)(\vec{\cal P} F\vec\pi)+$ \\

     &      & $\frac{g}{8}\omega^i\partial^i(SF)-\mu F^{0i}{\cal P}^{[0}\omega^{i]}]$  & $\frac{g}{8}\pi^i\partial^i(SF)-\frac{g}{2} F^{0i}{\cal P}^{[0}\pi^{i]}]$ \\

\hline

$\omega^\mu$   & $-\frac{eg}{2{\cal P}^0c}(F\omega)^\mu$  & $\frac{\omega^0eg}{2{\cal P}^0c}(F\omega)^\mu$  &  $-{\cal
P}^\mu+\frac{\pi^0eg}{2{\cal P}^0c}(F\omega)^\mu$ \\

\hline

$\pi^\mu$  & $-\frac{eg}{2{\cal P}^0c}(F\pi)^\mu$  &  ${\cal P}^\mu+\frac{\omega^0eg}{2{\cal P}^0c}(F\pi)^\mu$  &    $\frac{\pi^0eg}{2{\cal P}^0c}(F\pi)^\mu$   \\

\hline

$J^{\mu\nu}$  & $-\frac{eg}{2{\cal P}^0c}(FS)^{[\mu\nu]}$  & $\frac{\omega^0eg}{2{\cal P}^0c}(FS)^{[\mu\nu]}-2{\cal P}^{[\mu}\omega^{\nu]}$   & $\frac{\pi^0eg}{2{\cal P}^0c}(FS)^{[\mu\nu]}-2{\cal P}^{[\mu}\pi^{\nu]}$ \\

\hline
\end{tabular}
\end{center}
\end{table}
We will use the notation
\begin{eqnarray}\label{pht15.1}
u^0 &=& {\cal P}^0-\frac{(g-2)a}{2}(SF{\cal P})^0+\frac{ga}{8}S^{0\mu}\partial_\mu(FS), \qquad
a=\frac{-2e}{4m^2c^3-e(g+1)(SF)}, \cr \triangle^{\mu\nu} &=&v-\frac{2ca}{eu^0}{\cal P}^{(0}S^{\mu\nu)}, \qquad {\cal
P}^{(0}S^{\mu\nu)}={\cal P}^0S^{\mu\nu}+{\cal P}^\mu S^{\nu 0}+{\cal P}^\nu S^{0\mu}, \cr K^{\mu\nu}&=&
-\frac{gca}{4eu^0}S^{0\mu}\partial^\nu(SF), \qquad L^{\mu\nu\alpha} =-\frac{ga}{u^0}(FS)^{[\mu\nu]}S^{0\alpha}, \cr
g^{\mu\nu} &=& \eta^{\mu\nu}-\frac{2ca{\cal P}^0}{eu^0}{\cal P}^\mu{\cal P}^\nu. \qquad \qquad \qquad
\end{eqnarray}
Using the table, we obtain $\{T_3, T_4\}=\frac{eu^0}{2ca{\cal P}^0}$. Then
\begin{eqnarray}\label{pht14}
\{x^i, x^j\}_{D}=\frac12\triangle^{ij}, \qquad  \{x^i, {\cal
P}^j\}_{D}=\delta^{ij}-\frac{e}{2c}\left[\triangle^{ik}F^{kj}-K^{ij}\right], \nonumber
\end{eqnarray}
\begin{eqnarray}\label{pht14.1}
\{{\cal P}^i, {\cal
P}^j\}_{D}=\frac{e}{c}F^{ij}-\frac{e^2}{2c^2}\left[F^{ik}\triangle^{kn}F^{nj}-F^{[ik}K^{kj]}\right], \nonumber
\end{eqnarray}
\begin{eqnarray}\label{pht14.2}
\{S^{\mu\nu}, S^{\alpha\beta}\}_{D}= 2(g^{\mu\alpha} S^{\nu\beta}-g^{\mu\beta} S^{\nu\alpha}-g^{\nu\alpha} S^{\mu\beta}
+g^{\nu\beta} S^{\mu\alpha})+L^{\mu\nu[\alpha}{\cal P}^{\beta]}, \qquad \quad
\end{eqnarray}
\begin{eqnarray}\label{pht14.3}
\{S^{\mu\nu}, x^j\}_{D}={\cal P}^{[\mu}\triangle^{\nu]j}+\frac12 L^{\mu\nu j}, \nonumber
\end{eqnarray}
\begin{eqnarray}\label{pht14.4}
\{S^{\mu\nu}, {\cal P}^j\}_{D}=\frac{e}{c}\left[-{\cal P}^\mu(\triangle^{\nu k}F^{kj}-K^{\nu
j})-(\mu\leftrightarrow\nu)+\frac12 L^{\mu\nu k}F^{kj}\right]. \nonumber
\end{eqnarray}

\end{document}